# Automated Software Testing Starting from Static Analysis: Current State of the Art

Yan Wu[1], Jingyi Su[2,*], David D. Moran[3], and Chris D. Near[4]
[1,2]Bowling Green State University, Bowling Green, Ohio, United States
[3,4]Spectare-Systems, Inc., United States
yanwu@bgsu.edu, jsu@bgsu.edu, dmoran@spectare-sys.com, cdnear@spectare-sys.com

*Abstract*—The mass production of complex software has made it impossible to manually test it for security vulnerabilities. Automated security testing tools come in a variety of flavors, function at various stages of software development, and target different categories of software vulnerabilities. It is great that we have a plethora of automated tools to choose from, but it is a problem that their adoption and recognition are not prominent. The purpose of this study is to explore the possibilities of existing techniques while also broadening the horizon by exploring the future of security testing tools and related techniques.

## I. INTRODUCTION

Software is becoming increasingly important in many aspects of modern life, which leads to the allure of exploiting software vulnerabilities to acquire privileges such as confidential data and financial profits. Because the sheer bulk of today's software makes complete manual testing impractical, even for specific features, automated testing solutions (static, dynamic, hybrid, and so on) are in great demand. The idea of using automated software testing tools on software sounds productive and promising, however, the reality may not be so positive - developers and QA team members have a high tendency to abandon tools due to high false-positive rates reported, interruptions caused by running those tools, confusing descriptions, and so on. In fact, security testing tools typically focus on specific programming languages, categories of software vulnerabilities, as well as relatively high falsepositive and hidden false-negative rates, all of which contribute to a general distrust of existing security testing tools among software development stakeholders.

Speaking different languages separated people and sabotaged the project of Tower of Babel, speaking different languages in vulnerabilities description can do the same. To make sure people understand each other when communicating identified software vulnerabilities, Common Weakness Enumeration (CWE) has gone a long way from sporadic classification and categorization of reported software vulnerabilities to a gigantic community effort. In real life, major tool vendors usually have their own dictionary of rules" in the house to identify vulnerabilities, while the compliance requirement and urges from the market force them to map their rules to standardized rule sets. This process leaves quite a lot of blanks, mismatches, and chaos due to different understandings of the concepts and nature of specific vulnerabilities.

Common Vulnerabilities and Exposures (CVE) was developed before CWE, and it keeps the records of reported vulnerabilities and exposures for public reference. While compare to CWE, each CVE entry is an occurrence of a software fault which leads to security problems that could be exploited by a malicious attacker. For now, the CVE is maintaining more than 160k entries and they span proprietary and open source projects. The reported vulnerabilities are collected and reviewed to verify, then assigned a CVE ID. The disparity of available resources to identify and report between commercial and open-source project reports could contribute to the incompleteness of this repository.

The rest of this paper is structured as follows: In Section II, some related research is introduced. In Section III, the development opportunities are demonstrated. In Section IV, the preliminary idea of SATriage is described. In Section V, an overview and some future work are provided.

## II. LITERATURE REVIEW

As a keynote speaker on HASE 2015, Paul E. Black pointed out that different Static Analyzers are designed for different purposes and they usually don't report the same flaws". 12 wishes were raised for Static Analyzers in his talk, and they cover aspects from complex environment handling to support for more programming languages, etc. Black presents the position that although Static Analyzers will not be able to bring Utopia to us, continuous improvements in Static Analyzers still will be a great support to the software development community. Also as an experienced researcher from software assurance, Black et al pointed out that the directions to improve the impact of SA tools on software quality include decreasing the cost of use and increasing the benefit. Decreasing the cost involves activities like reducing false positives and integrating more transparently with the software development process. Increasing the benefit involves activities like finding higher level, subtle, and more critical bugs and helping programmers understand and fix the problem.

Felderer et al [1] conducted a thorough survey on various categories of security testing techniques that work at various stages of the software development lifecycle, ranging from Model-Based Security Testing on Requirements, Code-Based Testing, and Static Analysis (SA) on Development, to Security Regression Testing during maintenance. The authors also mentioned six criteria for selecting Security Testing Approaches, such as attack surface, application type, supporting technologies, and so on. They also used a three-tiered business application as a running example to explain how to choose appropriate security testing tools.

For Black et al, [2], the goal was to present a list of specific technical approaches that have the potential to make a dramatic difference in reducing vulnerabilities, by stopping them before they occur, by finding them before they are exploited or by reducing their impact". Three categories of information were discussed, and are technical approaches, measures and metrics, and non-technical approaches. Among them, five technical approaches include formal methods, system-level security, additive software analysis, more mature domain-specific software development framework, moving target defense (MTD) and automatic software diversity elaborated in detail, as potential developing areas.

In "A Few Billion Lines of Code Later," [3], Nunes et al proposed an approach to design a benchmark for the evaluation of SA tools (SATs) that detect vulnerabilities in web applications based on real-world workloads and different ranking metrics. To build the workloads, the authors collected vulnerable applications and labels them in terms of vulnerable and non-vulnerable lines of code, then assigned them to representative usage scenarios based on their software quality. To demonstrate the benchmark's feasibility, the authors evaluated five open-source PHP SATs to detect SQLi and XSS vulnerabilities in the workload composed of 134 WordPress plugins. The experimental results showed that the best tool changes from one scenario to another and also depends on the class of vulnerabilities being detected. Finally, the authors evaluated the ranking obtained using their instantiation with the same methodology and the OWASP's BSA benchmark.

Bessey et al [4] provide another perspective from one of the SAT vendors, Coverity. The authors shared their experience of commercializing Coverity as a product from the research project. They raised quite a few interesting points from a rare angle, which might be easily neglected by researchers. For example, the correct configuration and usage of a complex SAT might contribute quite much to the results, or, false positives could create a bad cycle from which the developers start to dislike SATs and even abandon them. Too much information is at least as bad as nothing at all.

Interestingly, Sadowski et al [5] provided experience from the other side - actual software users. Six main reasons why engineers do not always use SATs or ignore warnings were listed: not integrated; not actionable; not trustworthy; not manifest in practice; too expensive to fix; warning not understood. From the users' perspective, either any tool reports an issue incorrectly, or the developer did not understand an actual fault and took no action, the consequence is the same: the issue is considered to be an "effective false positive", if "developers did not take positive action after seeing the issue". On the contrary, the authors state that even if any tool "incorrectly reports an issue, but developers make the fix anyway to improve code readability or maintainability, that is not an effective false positive".

Although software testing plays an important role in the real-life development cycle, it can be a misunderstood task for some developers. People should think like an attacker in the vulnerability identification process. Potter and McGraw [6] suggest that in order to analyze risk, using a risk-based approach from the perspective of the system architecture and the attacker is valuable. Moreover, the baseline for riskbased security testing is highly dependent on expertise and experience.

In the study by Emanuelsson and Nilsson [7], SATs for detecting vulnerabilities were classified into four categories: String and pattern matching methods with syntactic pattern matching, unsound dataflow analysis with semantic information, sound dataflow analysis, and path and context-sensitive, and tools based on model checking techniques. Potential improvements include code volume, user-interactive use, high pricing, and more disjunction flaws than known ones. The evaluations should be utilized throughout the development process, but the needs and techniques vary depending on where you are in the process.

To differentiate features utilized for automatically capturing vulnerabilities, Sion et al [8] evaluated an existing catalog of 19 security design defects and software architectures with possible defects based on issue sets (e.g., CWE). A systematic data-flow model representation with detection rules is desired to maintain an organized detection of security design flaws by targeting and tracking items. By mapping security design flaw criteria to certain element types, these can be used to automatically detect design flaws in certain models.

The researchers Johnson et al [9] investigated possible reasons why SATs are not broadly used by software developers and proposed other expectations for these tools. When using SATs, developers desired more support for collaborative development, better integration of the tools into individual development phases, an intuitive display of flaws and explanation of flaws in detail, automatic suggestions for code fixes, and user-friendly interactive features to configure the tools.

Imtiaz et al [10] performed a study on stack overflow alerts and they found how the developers are making use of the SO alerts provided by SATs. In this study, the developers are more likely to control the SATs by filtering the alerts. The false positive is still a big challenge because of the mismatches and differences between the developers' domain knowledge and the reported results/alerts generated by the SATs. According to their research, SAT is expected to add user-friendly features to their tool for more convenient alert checking and item-specific customization.

Another interesting but generally overlooked topic is the issue of "false negatives" in software coding. Because of the agnostic nature of the number of false negatives, few researchers address this field, while those false negatives, if critical, could be a major threat to every software product. Thung et al [11] collected reported and fixed defects from selected open-source projects and ran three SATs on them to examine "false negatives" for those tools. The experiments are well designed, while the threats to validity are quite obvious, especially the limitations to reported and fixed defects, which might be a small subset of all defects, further the SATs are not naturally good at identifying certain categories of defects, such as specification errors.

## III. DEVELOPMENT OPPORTUNITY EXPLORATION

The software has made our lives easier, and countless people work in software-related industries. Security support should be provided constantly, academic communities should also pay more attention to future research in vulnerability detection. How to standardize a safe environment is a challenge for all practitioners. Along with the Software Development Lifecycle, it is vital to protect software artifacts against errors caused by carelessness, improper usage, unexpected interactions, and so on, at all stages. Next, we will discuss and explore some development opportunities in seven aspects.

### A. Privacy

A healthy software environment only allows limited operational access to users. When programs are functioning, it makes the program exposed to security attackers. Here is a paradox that in general, the more flexible and complex the functionality is, the more vulnerable the security of the way data is exchanged becomes. Data encryption can make these data exchange pipelines more secure. Thus, project encryption and data encryption are what we need to consider. It could be safer if programs can extract suspicious or even unreasonable parts directly from the encrypted data. Security problems could arise even during the security scanning procedure. Furthermore, open-source tools and third-party resources lack documentation and are more likely to be exposed to attackers due to infrequent updates. People often neglect the vulnerability of the development or implementation environments where the SATs run.

### B. Scalability

In order to make software development more efficient and to enable more complex functionalities, engineers often use a variety of tools for development and testing. Numerous engineers contribute to the same open-source program without guaranteeing that everyone has received enough security training. Because of that, it is suggested to focus on secure development, both the developers and management people be aware of the need for software security. Development environments that can provide security fixes in real-time are desired. The sensitivity and preference of such fixes should continuously fit individual needs. Any security analysis tool that is incorporated effortlessly with cutting-edge DevOp might be critical to its success.

There can be many uncontrollable factors in manual operations. If a portion of code is shared with many developers, it keeps generating hidden security issues from release to release. Every project version update brings complex and inherited issues. The inherited issues refer to the bugs that are not detected or completely fixed during the last patch or maintenance, and they may lead to other new bugs that may become security-related issues as a result of regular changes to the source code, with or without unfixed legacy code. Customized analysis for each project should improve the accuracy and coverage of security judgments.

### C. Hybrid

Within the hardware limitation, a combination of static and dynamic analysis is generally sufficient to maximize code security level without excessive resource consumption or missing too many flaws. Matching the flexible demands and applying mixed analysis methods are gaining prominence. But SA has a high false-positive rate, and the dynamic analysis approach, however, misses many possible vulnerabilities. Nowadays, compared to traditional methods, Machine learning (ML) has made it possible to rely less on experts' experience. Datasets that contain a large amount of valid and balanced cases for the analysis are strictly related to the model accuracy. With wellperformed ML methods, no longer will much manual work be consumed in redundant test analysis efforts.

Although community initiatives like Common Attack Pattern Enumeration and Classification (CAPEC) attempt to summarize attack patterns to identify paths, techniques, and so on, real intrusions are usually unique, unexpected, and use a variety of techniques and routes. To maintain the accuracy of existing tools, reports generated by SATs are intended to have overlaps and uncovered defects. Those undiscovered patterns must be considered in addition to the source of publicly reported patterns in order to improve the vulnerability detection rate. Learning from a large number of open-source projects that do not require domain expertise can be beneficial in reducing the effort required to refine the vulnerability catalog and database. With labeled defective functions extracted from a project, they are naturally built as input datasets and after pre-processing on the source code of different versions, it is also a solution to perfect identifying vulnerabilities. This stage of the learning process may be regarded as further ideas and a supplement to the previously merged defects. Because the majority of vulnerability detection techniques are designed at the function level, the capability of code dependencies is not well considered. Additional potential vulnerabilities might be detected by breaking down the code using an abstract syntax tree and transforming it into other levels of representation.

According to the various scanning results by different analysis tools, an integrated platform can obtain numerous security suggestions, but not all of them can be fixed within a limited time. Therefore, sorting and ranking the relevance and importance of these suspected vulnerabilities should be carefully performed. With a standardized development process, both the software development cycle and the vulnerability life cycle (How a flaw is created, exploited, and fixed.) can be evaluated.

### D. Balance

We used to depend on human experts to decide if a vulnerability has severe impacts on the program, but we may also rely on programs in the future to determine if this vulnerability needs to be taken care of with a higher priority. For example, if a case when a program can not correctly transfer the input as designed, and the transferred intermediate output will be used in other insecure cases, will be severe that will be determined by the machine. Developers use a variety of SATs, but each has its unique set of restrictions, coverage, and false-positive rates. For instance, some tool settings are for resource management, while others are for security engagement. The performance, resource cost, and configuration complexity vary. Either select the tool that best meets their needs, or obtain a summary of the findings given by several tools. People often focus on vulnerabilities with a high degree of severe consequences, but those issues only cover around half of the overall faults. Vulnerabilities that are reasonably simple to attack need more attention under the security concerns.

*E. Diversity*

While many analysis techniques employ domain experts to assess vulnerability patterns and severity, feedback from the owners and developers is critical in generating more project-specific analysis reports. Because people from varied backgrounds may wish to take charge of the general recommendations since either business users or technical users have different goals for the project, as well as different security needs.

Extracting insights from any project-related sources can contribute to finding new vulnerability patterns. For example, we can look for potential vulnerabilities in comments near a function in source code, published commit descriptions by any user on project management platforms, issues, and bugs with discussions posted by the developers, and even analyzed user-specific coding patterns and development preference from individual developers. Future work on detecting vulnerabilities combining should definitely include both programming languages and human languages analysis results, which is advantageous but can result in a more diverse security report.

*F. Backtracking*

There is a need for different sorts of vulnerability detection preferences for different types of software or operating systems. As the number of programming languages and platforms grows, so does the quantity and variety of faults. Investigating the relationships between old and new vulnerabilities for their connections and originals can contribute to detection performance. Since vulnerabilities may go unnoticed for a long time without impacting the current functionality. It is worth the effort to find a mechanism to map those possible flaws.

*G. Accuracy*

Seldom studied or even mentioned, the False Negative rate by nature should have been emphasized both in academia and among practitioners. Hidden, missing vulnerabilities are like time bombs, which may explode at any time. While the exact false-negative rate is agnostic for any non-native software program, which usually is complex enough to hide certain types of defects more frequently than others. From the other end of the spectrum, simple test cases provide yes or no clearcut answers while generally considered unrealistic so that the results may not be acceptable to practitioners. Field studies like Imtiaz et al [11] are enlightening while the strenuous manual work and extensive threats to validity limit the potential replicable research works or application to extended projects. One potential direction is to inject vulnerabilities to relatively simple, clean, medium-size software segments as semi-synthesized test cases to examine and estimate the false negatives. But the threat to validity still exists, considering the mainstream false negatives in design, and specification-related software vulnerabilities. From these negatives, formal methods, which lead to formal proof of traceability, and education and training, which bring in awareness and necessary skills, checks to designers, and developers, are great candidates to help reduce design, specification related false negatives.

## IV. SATRIAGE WITH EXPECTATIONS

Because SATs are widely used, business and technical users may have to try a few different available tools before finding the one that meets their needs the most. Some tools can be based on open-source projects, and they can be also based on cloud-based services. The development environment is not a limitation and it provides flexibility. Some are able to detect problems in the code while it is being written, the real-time support could be helpful in some cases. Those are popular features of software testing tools for detecting program flaws. According to the budget limitation, users can choose from free analysis tools to paid options. The report generated after the code scanning can usually assist engineers in detecting and prioritizing the possible vulnerabilities that need to be addressed. Most programming languages are supported by existing tools and the tools can be used at the same time or in different development phases. Since there is a variety of alternative SATs that developers and security professionals use, related users are recommended to use one or more tools with specific configurations to achieve the best outcome of vulnerability detection.

SATriage is a newly developed analytic tool that pioneers the automation of the Application Security Testing (AST) defect analysis and triage process and allows the developer to fully understand each defect and its context, importance, and risk factors. SATriage is an intelligent interactive platform that employs an innovative suite of weighting algorithms, as well as proprietary defect relationship data, to accurately reduce false positives, identify the most likely to be exploited software defects, and highlight the most serious security concerns specific to a given application. SATriage examines flaws rapidly depending on the context of an application and provides a level of urgency to each one. Because defect importance varies significantly depending on the application architecture, functionality, and usage patterns, context-sensitivity is crucial in triage to highlight the issues that are actually urgent to an application.

The SATriage process starts with the import of results from various ASTs, followed by normalization, correction, and merging of the results. Each tool has its own terminology and taxonomy, which must be consistent. According to CyberSagacity, more than 60 percent of reported problems are misclassified and/or misaligned among AST tools.

Once defect reports are imported, normalized, and corrected, SATriage performs a detailed analysis of each defect:

- Determine the likelihood and importance of every possible consequence of each defect.
- Perform false positive assessment, offering specific quick check validation information.
- Perform attack vector and ease-of-exploitation analyses to determine the likelihood of attack.
- Triage and prioritize results, based on the application context, severity, and likelihood of consequences, ease-of-exploitation, and confidence in the result.

SATriage prioritizes the defects in order of importance after completing a defect analysis.

SATriage provides all results in the form of a ranked ordered list. To determine the ranking results, SATriage utilizes a weighting mechanism based on estimated relative probabilities.

The most significant difference between SATriage's methodology and other DevOps tools is the distinction between probability and possibility. Other than SATriage, other tools rank defects based on possibility - if a defect has the potential to be severe, it must be severe. This results in a three to a five-tiered ranking system.

SATriage, on the other hand, employs a set of calculated relative probabilities. SATriage may rank individual defects in order by applying estimated relative probability. Typically, SATriage analysis will produce 300 - 500 different tiers of results.

SATriage, for example, considers the likelihood that a defect would result in a severe consequence and ranks appropriately. Many potentially severe defects are pushed down in the ranking because there is a low possibility that they will cause a severe consequence (along with a high probability of causing inconsequential issues). A tool using the possibilities method will discover that 30-50 percent of the defects are severe (i.e., 10s to 100s thousands of defects). The severity of the defects is counted in the dozens with the probability method, but more crucially, you receive an ordered list from 1 to N of which defects to address first, rather than a bucket of tens of thousands of defects with no differentiation in importance based on application context.

SATriage itself is under development and will be introduced in detail in another full paper.

## V. CONCLUSION

Software security is an area that needs a lot of attention but there are not enough people who are actually committed to supporting this work. Personal information is exposed to security attackers easily. Therefore, software security personnel has a great responsibility in guarding the privacy of the world. In this paper, we summarize and propose aspects that can contribute to the future security of software development, especially for SA software testing tools. How to balance security and convenience with a highly flexible software development process is still a challenge. Furthermore, technical and business users involved in any development phase should be sensitive during security development. We also briefly introduced SATriage and its development expectations - a more accurate and customizable tool for software vulnerability analysis that can be used in the future development cycle.

ACKNOWLEDGMENT

The authors express their gratitude for the research work supported by Spectare Systems, Inc., under contract to Bowling Green State University under contract reference.

REFERENCES

[1] Michael Felderer, Matthias Buchler, Martin Johns, Achim D. Brucker,¨ Ruth Breu, and Alexander Pretschner. "Security testing: A survey." Advances in Computers, vol. 101, pp. 1-51. Elsevier, 2016.
[2] Paul Black, Mark Badger, Barbara Guttman, and Elizabeth Fong. Dramatically reducing software vulnerabilities: Report to the white house office of science and technology policy. NIST Internal or Interagency Report (NISTIR) 8151 (Draft). National Institute of Standards and Technology, 2016.
[3] Paulo Nunes, Iberia Medeiros, Jos´e C. Fonseca, Nuno Neves, Miguel´ Correia, and Marco Vieira. "Benchmarking static analysis tools for web security." IEEE Transactions on Reliability 67, no. 3 (2018): 1159-1175.
[4] Al Bessey, Ken Block, Ben Chelf, Andy Chou, Bryan Fulton, Seth Hallem, Charles Henri-Gros, Asya Kamsky, Scott McPeak, and Dawson Engler. "A few billion lines of code later: using static analysis to find bugs in the real world." Communications of the ACM 53, no. 2 (2010): 66-75.
[5] Caitlin Sadowski, Edward Aftandilian, Alex Eagle, Liam Miller-Cushon, and Ciera Jaspan. "Lessons from building static analysis tools at google." Communications of the ACM 61, no. 4 (2018): 58-66.
[6] Bruce Potter, and Gary McGraw. "Software security testing." IEEE Security Privacy 2, no. 5 (2004): 81-85.
[7] Par Emanuelsson, and Ulf Nilsson. "A comparative study of industrial¨ static analysis tools." Electronic notes in theoretical computer science 217 (2008): 5-21.
[8] Laurens Sion, Katja Tuma, Riccardo Scandariato, Koen Yskout, and Wouter Joosen. "Towards automated security design flaw detection." In 2019 34th IEEE/ACM International Conference on Automated Software Engineering Workshop (ASEW), pp. 49-56. IEEE, 2019.
[9] Brittany Johnson, Yoonki Song, Emerson Murphy-Hill, and Robert Bowdidge. "Why don't software developers use static analysis tools to find bugs?." In 2013 35th International Conference on Software Engineering (ICSE), pp. 672-681. IEEE, 2013.
[10] Nasif Imtiaz, Akond Rahman, Effat Farhana, and Laurie Williams. "Challenges with responding to static analysis tool alerts." In 2019 IEEE/ACM 16th International Conference on Mining Software Repositories (MSR), pp. 245-249. IEEE, 2019.
[11] Ferdian Thung, David Lo, Lingxiao Jiang, Foyzur Rahman, and Premkumar T. Devanbu. "To what extent could we detect field defects? an empirical study of false negatives in static bug finding tools." In 2012 Proceedings of the 27th IEEE/ACM International Conference on Automated Software Engineering, pp. 50-59. IEEE, 2012.